\documentclass[12pt]{article}

\usepackage{newtxtext,newtxmath}

\usepackage{graphicx} 
\usepackage{color} 
\usepackage[T1]{fontenc}
\usepackage[letterpaper,margin=1in]{geometry}

\renewenvironment{abstract}
	{\quotation}
	{\endquotation}

\date{}

\makeatletter
\renewcommand{\fnum@figure}{\textbf{Figure \thefigure}}
\renewcommand{\fnum@table}{\textbf{Table \thetable}}
\makeatother

\usepackage{scicite}

\usepackage{url}

\def\scititle{
	Non-maximal entanglement of photons from positron-electron annihilation demonstrated using a plastic PET scanner
}
\title{\bfseries \boldmath \scititle}

\author{
    Pawel~Moskal$^{1,2,3\ast}$,
    Deepak~Kumar$^{1,2,3}$,
    Sushil~Sharma$^{1,2,3\ast\ast}$,\and
    Ermias Yitayew Beyene$^{1,2,3}$,
    Neha~Chug$^{1,2,3}$,
    Aurélien~Coussat$^{1,2,3}$,\and
    Catalina~Curceanu$^{4}$,
    Eryk~Czerwi{\'n}ski$^{1,2,3}$,
    Manish~Das$^{1,2,3}$,
    Kamil~Dulski$^{1,2,3}$,\and
    Marek~Gorgol$^{5}$,
    Bozena~Jasi{\'n}ska$^{5}$,
    Krzysztof~Kacprzak$^{1,2,3}$,\and
    Tevfik~Kaplanoglu$^{1,2,3}$,
    {\L}ukasz~Kaplon$^{1,2,3}$,
    Tomasz~Kozik$^{1,2,3}$,\and
    Edward~Lisowski$^{6}$,
    Filip~Lisowski$^{6}$,
    Wiktor~Mryka$^{1,2,3}$,
    Szymon~Nied{\'z}wiecki$^{1,2,3}$,\and
    Szymon~Parzych$^{1,2,3}$,
    Elena P.~del Rio$^{1,2,3}$,
    Martin~R{\"a}dler$^{1,2,3}$,\and
    Magda~Skurzok$^{1,2,3}$,
    Ewa~{\L{}}.~Stepie{\'n}$^{1,2,3}$,
    Pooja~Tanty$^{1,2,3}$,\and
    Keyvan~Tayefi~Ardebili$^{1,2,3}$ and
    Kavya~Valsan~Eliyan$^{1,2,3}$ \and
	\small$^{1}$Faculty of Physics, Astronomy and Applied Computer Science, Jagiellonian University, \and \small S.~Łojasiewicza 11, 30-348 Kraków, Poland.\and
	\small$^{2}$Total-Body Jagiellonian-PET Laboratory, Jagiellonian University, Poland.\and
    \small$^{3}$Center for Theranostics, Jagiellonian University, 31-034 Kraków, Poland.\and
    \small$^{4}$INFN, Laboratori Nazionali di Frascati CP 13,  Via E. Fermi 40, 00044, Frascati, Italy.\and
    \small$^{5}$Institute of Physics, Maria Curie-Sklodowska University, \and\small Pl.~M.~Curie-Sklodowskiej~1, 20-031 Lublin, Poland.\and
   \small$^{6}$Cracow University of Technology, 31-864 Kraków, Poland.\and
	\small$^\ast$Corresponding author. Email: p.moskal@uj.edu.pl\and
	\small$^{\ast\ast}$Corresponding author. Email: sushil.sharma@uj.edu.pl
}

\begin{document} 


\maketitle
\begin{abstract} \bfseries \boldmath
{
In state-of-the-art Positron Emission Tomography (PET), information about annihilation photon polarization is unavailable. Here, we present a PET scanner built from plastic scintillators, where annihilation photons primarily interact via the Compton effect, providing information about both photon polarization and propagation direction. Using this plastic-based PET, we determined the distribution of the relative angle between polarization planes of photons from positron-electron annihilation in a porous polymer. The amplitude of the observed distribution is smaller than predicted for maximally quantum-entangled two-photon states but larger than expected for separable photons. This result can be well explained by assuming that photons from pick-off annihilation are not entangled, while photons from direct and para-positronium annihilations are maximally entangled. Our result indicates that the degree of entanglement depends on the annihilation mechanism in matter, opening new avenues for exploring polarization correlations in PET as a diagnostic indicator.
}

\end{abstract}


\section*{\textbf{Introduction}}\label{sec:introduction}
\begin{figure}[!h]
    \centering
    \includegraphics[scale=.4]{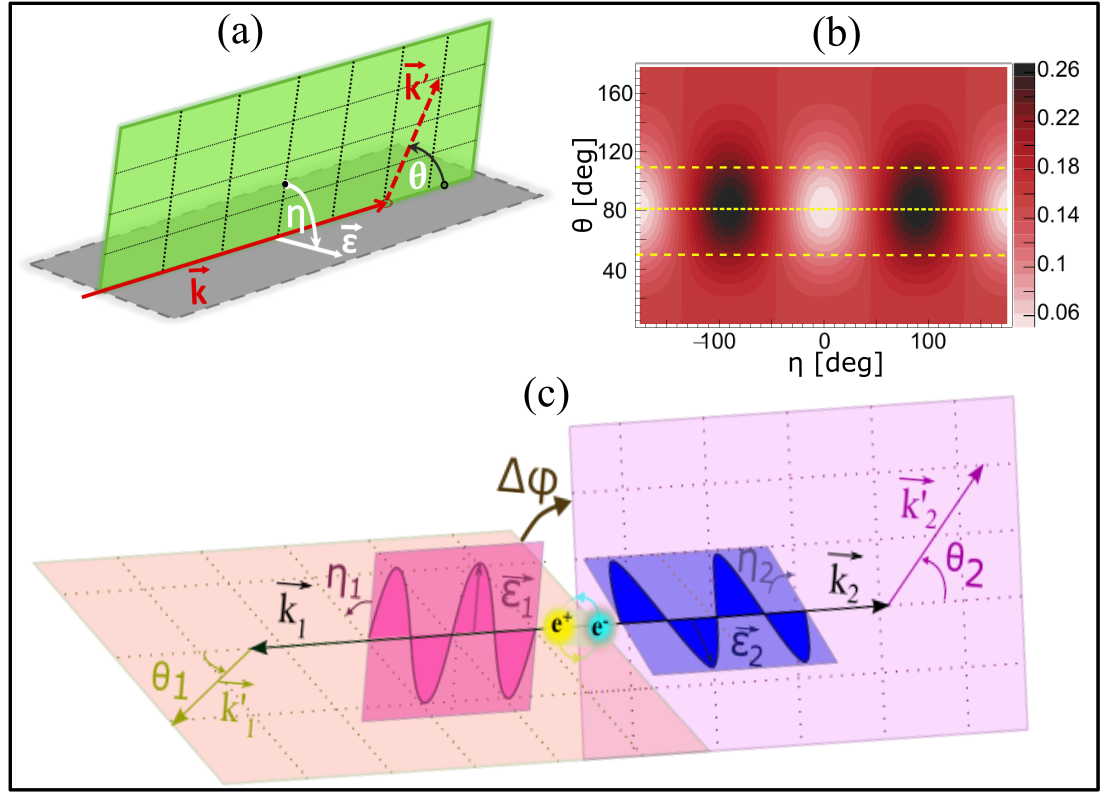}
    \caption{
    {
        {\bf{Relation between scattering angle and photon polarization in the Compton scattering of annihilation photons.}}
    }  
    \textbf{(a)} Pictorial illustration of the photon-electron Compton scattering with the definition of the photon scattering angle ($\theta$), photon polarization ($\vec{\varepsilon}$), photon momentum before ($\vec{k}$) and after ($\vec{k'}$) the scattering, and the angle ($\eta$) between the polarization and scattering planes.
    \textbf{(b)} 
    Normalized Klein-Nishina cross section for 511 keV photons as a function of angles $\eta$ (horizontal axis) and $\theta$ (vertical axis). For normalization, the 2D plot ($\eta$, $\theta$) was weighted such that for each value of $\theta$ the integral of cross-section over the entire range of $\eta$ from $-\pi$ to $\pi$ is equal to unity. The yellow dotted line indicates the value of $\theta$~=~82$^\circ$, for which visibility is maximum~\cite{Beatrix:2019} and dashed lines indicate a range of high visibility for 
    $\theta~=~82^\circ~\pm~30^\circ$~\cite{Moskal:2018}.
    \textbf{(c)} 
    Schematic representation of annihilation photons and their Compton scattering, including polarization and scattering planes. $\theta_1$ and $\theta_2$ denote the Compton scattering angles. $\eta_1$ and $\eta_2$ denote angles between the scattering and polarization planes. $\Delta\varphi$ represents the angle between the scattering planes of annihilation photons and thus is a measure of the relative angle between their polarization planes.}
    \label{scheme}
\end{figure}

Positron emission tomography (PET) is an established imaging technique for non-invasive {\it {in vivo}} diagnosis of disease in clinical practice ~\cite{Schwenck:2023,Richard:2014,Gustav:2006}. 
Its potential for quantitative assessment of metabolic alteration in biological tissues makes it useful for various medical applications by assessing physiology (functionality) of human organs or tissues~\cite{Gambhir:2012}. In PET a biomolecular tracer labeled with a positron ($e^+$) emitting radionuclide is administered into the human body. The emitted $e^+$ interacts with electron ($e^-$) in the tissue and annihilates predominantly into two 511 keV photons moving in opposite directions. 
The principle of PET is based on the registration of places and times of interactions of these two photons, and reconstruction of the site of annihilation along their direction of propagation   referred to as Line of Response (LOR).
The information from the LORs is used as input to reconstruct the density distribution of annihilation points. 
However, annihilation photons carry more information than just about the site where they originated. Generally, annihilation photons carry information in the form of energy, direction of propagation, polarization, and the degree of entanglement~\cite{McNamara:2014,Toghyani:2016,Romanchek:2023}. Polarization of annihilation photons is not accessible by current PET systems, but in principle it can inform us about the contributing annihilation mechanisms, that in turn may tell us about the cell molecular composition. In the body, a positron emitted from the isotope attached to the biomolecule may annihilate with an electron either directly or via formation of positronium ~\cite{NatureReviewPhysics:2019,Bass:2023rmp}. Positronium in the tissue intermediates the positron-electron annihilation in about 40\% of cases~\cite{Harpen:2004,Jasinska:2017b,Moskal:2019a}. Positronium may be formed as a long lived (142 ns) spin-one ortho-positronium (oPs), or as a short lived (125 ps) spin-zero parapositronium (pPs)~\cite{CASSIDY:2018}. In vacuum oPs decays into 3-photons (oPs → 3$\gamma$) and pPs into 2-photons (pPs → 2$\gamma$)~\cite{Bass:2023rmp}. Theoretically, photons from the decay of positronium in vacuum are maximally entangled in polarization~\cite{Acin:2001,Beatrix:2019,Nowakowski:2017}. However, in matter when positron from positronium annihilates with the electron bound to the atom, it is natural to ask the question of whether the photons resulting from such annihilation are maximally entangled
~\cite{Coffman:2000,Koashi:2004,Horodecki:2009,Caradonna:2024}. 
Annihilation photons possess energy in the range of MeV and hence interact in matter with single electrons. Therefore, their polarization cannot be studied using optical methods. However, polarization of such high energetic photons can be estimated by Compton scattering (Fig.~\ref{scheme}(a)). The Compton scattering of photons is most likely in a plane perpendicular to the polarization of the incoming photon~\cite{KleinNishina:1929}, and therefore polarization orientation of the primary photon ($\vec{\varepsilon}$) at the moment of scattering can approximately be determined as a vector product of momentum vectors of initial $\vec{k}$ and scattered photon $\vec{k'}$ ~\cite{Moskal:2016}:
\begin{equation}\label{polarisation}
    \vec{\varepsilon} = \frac{\vec{k} \times \vec{k'}}{ |\vec{k_{i}}\times \vec{k_{i}^{{}'}}|}
\end{equation}

 Fig.~\ref{scheme}(b) describes the distribution of $\eta$~\cite{Moskal:2018}, the relative angle between polarization plane and scattering plane, as a function of the scattering angle $\theta$. It indicates that, in case of 511~keV annihilation photons, the correlation between the polarization direction $\vec{\varepsilon}$ and the scattering plane is maximal at about $\theta$ = $82^\circ$  decreasing to negligible effect for forward ($\theta$ = $0^\circ$) and backward ($\theta$ = $180^\circ$) scattering. In general the analysing power (A$_p$) of the Compton polarimeter is expressed as ~\cite{Beatrix:2019,Ivashkin:2023,Knights:2018}:
\begin{equation}
A_p~=~sin^2\theta / (E/E' + E'/E - sin^2\theta),    
\end{equation}
where E and E$'$ denotes the energy of primary and scattered photon, respectively~\cite{Knights:2018}. For 511~keV photons, maximum value of A$_p$ equal to 0.69 is reached at $\theta$~=~82$^\circ$.

The distribution shown in Fig.~\ref{scheme}(b) indicates that the Compton scattering may be effectively applied as polarimeter in the scattering angle range of about $\theta$ = $82^\circ$ $\pm$ 30$^\circ$, in which analysing power varies from 0.69 to 0.42. This range is indicated by horizontal dashed lines.
For the 2$\gamma$ annihilation process (Fig.~\ref{scheme}(c)), when each $\gamma$ interacts via Compton scattering with an electron, one can estimate the relative angle between the polarization directions of the photons $|\eta_1$ - $\eta_2|$ by measurement of the relative angle $\Delta\varphi$ 
between the scattering planes~\cite{Moskal:2018}. Bose-symmetry and parity conservation in the decay of parapositronium (pPs) imply that the state $|\psi\rangle$ of the resulting two photons is maximally entangled and that the photons polarizations are orthogonal to each other~\cite{Beatrix:2019}. In the linear polarization base the 2-photon state from pPs decay reads:
\begin{equation}
\centering
|\psi\rangle=\frac{1}{\sqrt{2}}(|H\rangle_{1}|V\rangle_{2}+|V\rangle_{1}|H\rangle_{2}),
\end{equation}
where H and V correspond to the horizontal and vertical polarization directions. Taking into account the Compton scattering of each of the photons, 
the double Compton scattering differential cross section can be expressed as~\cite{Pryce:1947}: 
\begin{equation}\label{DDx2}
\begin{aligned}[b]
  \frac{d^{2}\sigma(\theta_1,\theta_2,\Delta\varphi)}{d\Omega_{1}d\Omega_{2}}= &\left[A(\theta_{1},\theta_{2})-B(\theta_{1},\theta_{2})cos(2(\Delta\varphi))\right],
\end{aligned}
\end{equation}
where, A and B describe dependence of the cross section on the Compton scattering angles $\theta_{1}$ and $\theta_{2}$. At given scattering angles ($\theta_{1}$,$\theta_{2}$), the cross section is maximal for $\Delta\varphi$ = 90$^{\circ}$ and minimum for $\Delta\varphi$ = 0$^{\circ}$. 
The strength of the polarization correlation of the annihilation photons may be described by R factor (referred to as entanglement witness), which is the ratio of the probabilities of the scattering at $\Delta\varphi$ = 90$^{\circ}$ and $\Delta\varphi$ = 0$^{\circ}$. This implies:  
\begin{equation}\label{EW}
\begin{aligned}[b]
 R = \frac{A(\theta_{1},\theta_{2})+B(\theta_{1},\theta_{2})}{A(\theta_{1},\theta_{2})-B(\theta_{1},\theta_{2})},
\end{aligned}
\end{equation}
The maximum of the cross section at $\Delta\varphi$ = 90$^{\circ}$ reflects the fact that the polarizations of the photons are perpendicular to each other. For maximally entangled photons, the value of R reaches a maximum of 
{R$_{max}$~=}~2.84 at the scattering angles of 
$\theta_1 = \theta_2 = 82^{\circ}$~\cite{Beatrix:2019,Pryce:1947,Snyder:1948, Hanna:1948, Wu:1950, Langhoff:1960, Kasday:1975, Watts:2021, Bohm:1957,Parashari:2022}, while for separable photons the maximum value of R is equal to 
{R$_{sep}$~=}~1.63~\cite{Watts:2021,Bohm:1957}. The shape of the distribution of the angle $\Delta\varphi$ can in principle carry information about the molecular composition of annihilation sites. In matter, 2$\gamma$ annihilations (used in PET) occur either by (i) direct electron-positron annihilation, (ii) via self-annihilation of pPs or (iii) via oPs annhilation due to interaction with surrounding electrons. The oPs in addition to self-annihilation into 3-photons, may annihilate into 2-photons when the positron from orthopositronium picks off electron from the surrounding molecular environment (pick-off process)~\cite{Bass:2023rmp}. Here we hypothesize that the $\Delta\varphi$ distribution may depend on the annihilation mechanism.
{
In general, we assume that in an $\alpha$ fraction of 2$\gamma$ annihilations, the photons are maximally entangled and in the remaining (1-$\alpha$) fraction the produced photons propagate independently of each other.  Thus, without loss of generality the measured $\Delta\varphi$-distribution F($\Delta\varphi$) may be decomposed as 
\begin{equation}
F(\Delta\varphi,\alpha) = \alpha F(\Delta\varphi,R_{max}) + (1-\alpha) F(\Delta\varphi,R_{sep}). 
\end{equation}
Using this formula one can show that:
\begin{equation}
 R(\alpha) = \frac{( 1 + \alpha (R_{max}-1)/(R_{max}+1) + (1-\alpha)(R_{sep}-1)/(R_{sep}+1) )}{
                 (  1 - \alpha (R_{max}-1)/(R_{max}+1) - (1-\alpha)(R_{sep}-1)/(R_{sep}+1) )}.
\end{equation} 
Trying to estimate an expected effect of R variation in the material, we assume further that photons from the pick-off annihilations are separable (R = R$_{sep}$), and photons from other processes (direct annihilation, pPs decay and oPs conversion) are maximally entangled (R = R$_{max}$). 
The assumption about maximal entanglement for direct annihilation is justified since previous investigations of the $\Delta\varphi$ distribution for annihilation in metals are consistent with the predictions for maximally entangled photons~\cite{Ivashkin:2023}, and in metals, positrons annihilate only directly with electrons. Also, self-annihilation of pPs, theoretically leads to the pair of maximally quantum entangled photons~\cite{Acin:2001}. However, in the case when oPs annihilates via interactions with an electron from the surrounding atoms (pick-off process), the initial state will be a mixture of electron-positron states with many possible quantum numbers, and therefore it is plausible to assume that in this case for the mixed and not a pure quantum state, the resulting photons will not be entangled. 
}
This would mean that value of R depends on the material and may serve as an indicator of the intra-molecular environment surrounding the positronium.
%
%
In this work, we present a dedicated positron emission tomography scanner, called the Jagiellonian PET (J-PET), built from plastic scintillators in which annihilation photons interact solely via Compton scattering (with photoelectric effect at the level of 10$^{-5}$~\cite{Moskal:2021pmb}). In the J-PET scanner, the average distance between the primary and secondary scattering of 511 keV photon is equal to about 25~cm (compared to about 0.6~cm in crystal PET systems)
rendering possible measurement of the $\Delta\varphi$ distribution with high angular resolution {of 2$^\circ$}~\cite{Moskal:2018}. 
We substantiate the capability of the J-PET scanner in effectively imaging the $\Delta\varphi$ distribution and hence for imaging of parameter R in addition to the standard PET imaging of the density distribution of annihilation points. As an example of application we  determine that value of R for photons from the positron-electron annihilation in porous polymer is significantly lower then expected for maximally entangled photons, and by comparison to the value of R determined for aluminum, we demonstrate that R is sensitive to the type of the material. This result opens up prospects for using entanglement witness R as a diagnostic parameter of tissue type and tissue pathology. 
Finally, we estimate and discuss the sensitivity of the total-body J-PET scanner for the simultaneous PET and R-value imaging in clinical diagnostics.\\

\section*{Results}\label{sec:results}

\begin{figure}[!h]
    \centering
    \includegraphics[scale = .2]{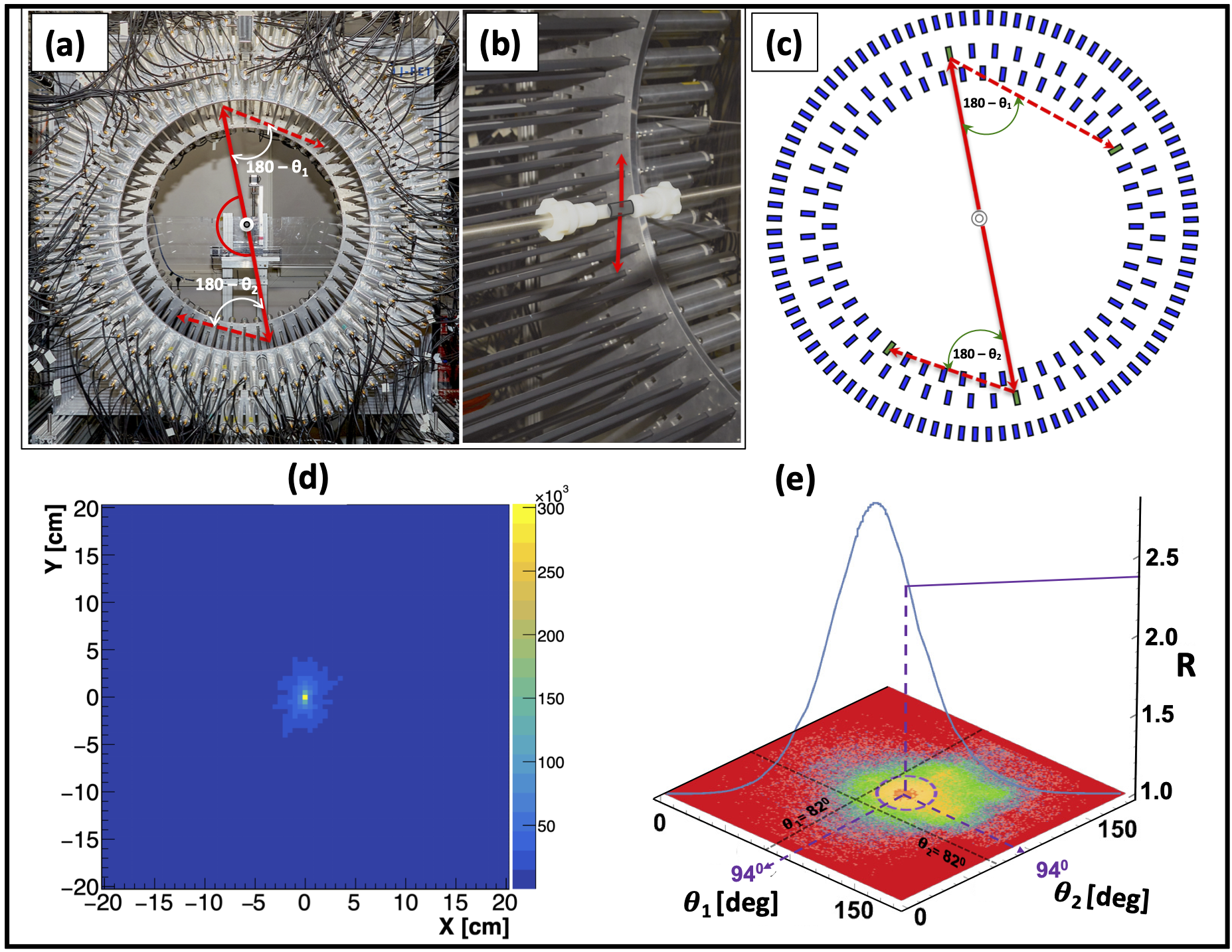}
    \caption{
        {
        {\bf{Experimental setup used in this study and exemplary distributions of registered events.}}
    }
    \textbf{(a)} Photograph of the 192-strip J-PET tomograph with a superimposed illustration of the example event with annihilation photons (solid red arrows) and scattered photons (dashed red arrows) used in this study.  
    \textbf{(b)} Close-up photograph of the annihilation chamber, which comprises a positron-emitting $^{22}$Na radionuclide surrounded by XAD-4 porous polymer~\cite{Jasinska:2016}. In this photograph, scintillator strips covered with black light-tight foil and aluminum photomultiplier housings are visible.
    \textbf{(c)} The cross-section of the 192-strip J-PET scanner together with the annihilation photons (solid red lines) originating from the electron-positron annihilation in the chamber, and the scattered photons (dashed red lines).
    \textbf{(d)} Tomographic image of the annihilation source with a pixel size of 5~mm~$\times$~5~mm.
    \textbf{(e)} Experimentally determined distribution of scattering angles $\theta_{2}$ vs. $\theta_{1}$. The maximum density of events determined at $\theta_{1}~=~\theta_{2}~=~94^\circ$ is due to the geometry of the detector, as can be seen in Figs. (a) and (b). The superimposed solid-blue curve indicates value of entanglement witness R calculated for the cases where $\theta_{1}~=~\theta_{2}$. The maximum of R~=~2.84 is visible for $\theta_{1}~=~\theta_{2}~=82^\circ$, while the experimental data concentrates around $\theta_{1}~=~\theta_{2}~=~94^\circ$ where R~=~2.4.}
    \label{fig:2}
\end{figure}

This work experimentally demonstrates for the first time the dependence of the degree of entanglement of annihilation photons on the type of the material in which the positron annihilates, and presents the capability of newly developed J-PET scanner to image the quantum correlation of annihilation photons. Figs.~\ref{fig:2}(a) and~\ref{fig:2}(b) show photographs of the J-PET scanner developed and constructed by 
the J-PET group~\cite{Moskal:2021pmb,Moskal:2014nim,Niedzwiecki:2017acta,Sharma:2020}. The technical details of the J-PET scanner and the annihilation chamber are outlined in the Methods section. In this section, we highlight the key features that set J-PET apart from the current PET scanners and make it capable of imaging quantum entanglement of annihilation photons. The J-PET scanner is built from plastic scintillator strips arranged axially with the photomultiplier readout at the ends. Application of plastic scintillators, instead of crystals used in the state-of-the-art PET systems, and the application of dedicated triggerless data acquisition system are the two crucial novelties enabling efficient detection of events in which both photons from $e^+e^- \to 2\gamma$ annihilation undergoes Compton scattering. In plastic scintillators 511~keV annihilation photons interact via Compton scattering only (fraction of photoelectric effect is at the order of 10$^{-5}$~\cite{Moskal:2021pmb}), and triggerless acquisition enables simultaneous detection of 4 interactions due to annihilation and scattered photons. Contrary to the state-of-the-art PET systems in which signal processing and acquisition are confined to two interactions only~\cite{Jones:2017,Surti:IEEE2020}. The registration of Compton scattered photons is mandatory for the determination of the annihilation photon's polarization. 

The linear polarization of the i$^{th}$ annihilation photon is determined using equation \eqref{polarisation} ~\cite{Moskal:2016,Moskal:2018}.
A typical topology of events used in this study is superimposed on the photograph of the scanner (Fig.~\ref{fig:2}(a)),  and on its schematic cross section (Fig.~\ref{fig:2}(c)). Positron-electron annihilation occurs in the porous polymer XAD-4~\cite{Jasinska:2016} surrounding the positron emitting $^{22}$Na source placed in annihilation chamber Fig.\ref{fig:2}(b). The two annihilation photons are emitted in opposite directions and registered in the three-layer J-PET system consisting of 192 plastic scintillators of 50~cm length and a cross section of 1.9$\times$0.7~cm$^2$~\cite{Niedzwiecki:2017acta}. The signals from photomultipliers are sampled by dedicated electronics~\cite{Palka:2017wms} enabling determination of time, position and energy deposition for each registered photon interaction~\cite{Niedzwiecki:2017acta}. Interactions are referred to as hits.  For an event of interest, four hits are required, two hits from annihilation photons and two hits from scattered photons. Hits in the detector caused by the annihilation photons are distinguished from hits due to scattered photons based on the registered energy deposition and angular correlations. Hits from annihilation photons are used to reconstruct the tomographic image of the annihilation source. An example image, showing the reconstructed position of the source, is presented in Fig.~\ref{fig:2}(d). Next, each scattered photon is assigned to the appropriate annihilation photon using information on time and positions of hits. The events selection criteria are discussed in detail in the Methods section. 

Once the hits in the event are identified, the scatterings angles ($\theta_1$ and $\theta_2$) and the angle between the scattering planes ($\Delta\varphi$) are determined. The determined distribution of the $\theta_1$ versus $\theta_2$ angles is shown in Fig.~\ref{fig:2}(e). It illustrates that the maximum of the density distribution of registered events is around the angles of $\theta_1 = \theta_2 = 94^\circ$. This is due to the geometric arrangements of scintillators in the J-PET tomograph, as can be seen in Fig.~\ref{fig:2}(c). Solid curve superimposed on the plot in Fig.~\ref{fig:2}(e) indicate the distribution of entanglement witness R for scatterings where $\theta_1 = \theta_2$, calculated under the assumption that the annhilation photons are maximally entangled. 

\begin{figure}
    \centering
   \includegraphics[scale=0.48]{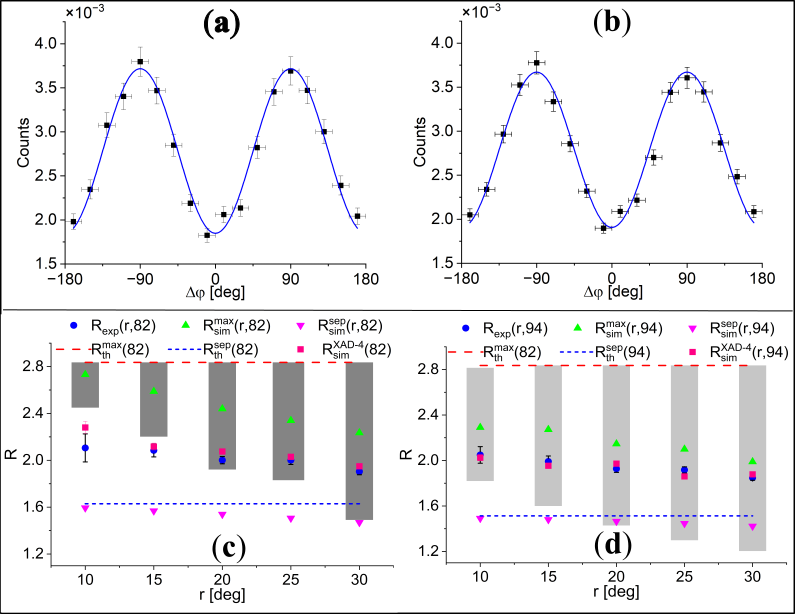}
    \caption{
{     
 {\bf{ The distributions of $\Delta\varphi$ angle and the values of R-factor determined for the XAD-4 porous material. }}   
 }
\textbf{(a,b)} The (a) and  (b) panels show experimental results for events within a circle with the radius of $20^\circ$ centered around $\theta_1 = \theta_2 = 82^\circ$, and $\theta_1 = \theta_2 = 94^\circ$, respectively. The solid curve represents the best fit of the function A-B$\cos(2\Delta\varphi)$, with A and B as free parameters. Vertical bars denote statistical uncertainty, while horizontal bars indicate bin width. The tabulated data are provided in Supplementary Tables~S1 and~S2. 
\textbf{(c,d)} R values determined for events within circles with the radii of r, centered around $\theta_1 = \theta_2 = 82^\circ$, and  $\theta_1 = \theta_2 = 94^\circ$, respectively (see also Fig.~\ref{fig:6}). Experimental results are shown with blue points. Black bars denote the statistical uncertainty. The maximum possible value of R for the case when photons are maximally entangled (R~=~2.84), achievable at $\theta_1 = \theta_2 = 82^\circ$, is indicated by the horizontal dashed-red line. The dashed-blue lines indicate the values of R in the case when the photons are not entangled for the scatterings at $\theta_1 = \theta_2 = 82^\circ$ (R~=~1.63) and  $\theta_1 = \theta_2 = 94^\circ$ (R~=~1.51) in (c) and (b) panels, respectively. Upward-pointing green triangles and downward-pointing magenta triangles demonstrate R values determined from the simulated data. Upward-pointing green triangles correspond to the result simulated assuming maximally entangled photons, while downward-pointing magenta triangles indicate results obtained assuming that the photons are separable. The pink rectangles correspond to simulated results for porous medium XAD-4 assuming that photons from the pick-off annihilation are not entangled. The shaded boxes show the range of theoretical R values, in a given selected circle, calculated for the case of maximally entangled photons.  
    \label{fig:3}
    }
\end{figure}

It shows that the value of R expected at maximum of the density distribution of registered events is equal to R~=~2.4, and it is lower than that the maximal value of R~=~2.84 for $\theta_1 = \theta_2 \approx 82^\circ$. 
For further analysis of the experimental data, we selected the region around $\theta_1 = \theta_2 = 82^\circ$ (112943 events), where the highest correlation is expected, and the region around
$\theta_1 = \theta_2 = 94^\circ$ (181186 events), where the highest number of events is recorded. Fig.~\ref{fig:3}~(a) and Fig.~\ref{fig:3}~(b) show the $\Delta\varphi$ distribution determined for events with the
$\theta_1,\theta_2$ scattering angles from the circle of radius r~=~20$^{\circ}$ centered at 
$\theta_1 = \theta_2 = 82^\circ$ (Fig.~\ref{fig:3}(a)) and at $\theta_1 = \theta_2 = 94^\circ$ (Fig.~\ref{fig:3}(b)), respectively. The studied regions on the ($\theta_1,\theta_2$) plot are shown in the Methods section.
The formula $A-B\cos(2\Delta\varphi$) was fitted to the determined distributions in order to extract the value of 
R$_{exp} = (A + B)/(A - B)$.
The blue curves depict the results of the fits superimposed on the data, and the resultant values of R$_{exp}$ are represented by blue circles in Fig.~\ref{fig:3}(c) and Fig.~\ref{fig:3}(d). 
The experimental values of R$_{exp}$ that were obtained correspond to the weighted average of R values, where the weights are based on the density distribution of registered events over the selected region on the ($\theta_1,\theta_2$) distribution.
In the figure, the experimental results are compared to the theoretical predictions obtained under two assumptions: (i) that the photons are maximally entangled (green triangles pointing upwards) and (ii) that the photons are in a separable state (magenta triangles pointing downwards). The shown theoretical predictions account for the effects of the detection system, which were modeled using Monte Carlo simulation methods as described in detail in the Methods section. The measured experimental values of the entanglement witness R$_{exp}$ (given in Table I) are smaller than the predicted values for maximally entangled photons R$^{max}_{sim}$, and larger than the expected values for separable photons R$^{sep}_{sim}$. The difference between the experimental and theoretical values of the entanglement witness R remains consistent across all studied regions with radius r ranging from r~=~10$^\circ$ to r~=~30$^\circ$. As the radius increases, the experimental value of R$_{exp}$ decreases, and the statistical errors also decrease. For the mean radius of r~=~20$^\circ$, the determined values of R$_{exp}$ for the distribution of scattering angles centered at $\theta_{1} = \theta_{2} = 82^{\circ}$ and $\theta_{1} = \theta_{2} = 94^{\circ}$ are estimated to be 2.00~$\pm$~0.03 and 1.93~$\pm$~0.03, respectively. 

\section*{{Discussions}}\label{sec:discussion}

This study demonstrates the first full-scale PET scanner capable of determining the polarization of annihilation photons by registering their Compton scatterings. The scanner is constructed from three cylindrical layers of plastic scintillator strips, in which annihilation photons interact almost exclusively through Compton interaction, compared to crystals where Compton effect constitutes e.g. 59\% (BGO) or 69\% (LYSO)
~\cite{NIST:PhotonDataBase,Moskal:2021pmb,Eijk:2003}. 
Therefore, plastic scintillators are the best suited for the quantum entanglement PET system. Another crucial characteristics of the scanner presented in this work is the large relative distance between subsequent Compton interactions in plastic scintillators. In the J-PET system it is equal on the average to 25~cm, compared to about 0.6~cm~in BGO crystal scintillators.
This enables one to achieve with J-PET a high angular resolution of about 2$^{\circ}$, and high purity of up to $\sim$95\%~\cite{Moskal:2018},
for the identification of first and second interactions, compared to only about 55\% purity and about 6$^\circ$ angular resolution achieved in pixelated crystals~\cite{Parashari:2022,Ana:2021}. Moreover, in the presented J-PET system, the maximum efficiency for detecting double Compton interactions is within the angular range of $82^\circ\pm30^\circ$, where the correlation is the highest (see Fig.~\ref{fig:2}(d)). 
Using the data collected with the J-PET scanner we determined the distributions of the relative angle between the scattering planes of 511~keV photons ($\Delta\varphi$) originating from $e^+e^-\to 2\gamma$ annihilation in the porous polymer XAD-4. The obtained shape of the $\Delta\varphi$ distributions exhibits cos(2$\Delta\varphi$) oscillations with a maximum at 90$^\circ$, as expected for photons with perpendicular polarizations. The experimental $\Delta\varphi$ distributions were compared with predictions obtained under the assumptions that the photons are maximally entangled and that they are separable. The main observation reported in this work is that the correlations between the annihilation photons originating from the positron-electron annihilation in the porous polymer are larger than for the separable state but they are smaller than expected for the maximally entangled two-photon state. This finding is reported for the first time. Previous investigations of the $\Delta\varphi$ distribution for annihilation in metals, resulted in the R-value consistent with the assumption that annihilation photons from the electron-positron annihilation are maximally entangled.\cite{Bordes:2023,Parashari:2024} 
The most precise experiment so far, performed for photons from positron-electron annihilation in aluminium, yielded 
R$ = 2.435 \pm 0.018$~\cite{Ivashkin:2023}, which taking into account the detector geometrical acceptance, is consistent with expectations for the maximally entangled two-photon state~\cite{Ivashkin:2023}. 
\begin{table}
    \centering
    \caption{{\bf{Parameter R$_{exp}$ determined for the range of scattering angles centered around 82$^\circ$ and 94$^\circ$ with the radius r~=~20~cm}}. The values are compared to the {{simulations of experimental result}} assuming that photons are maximally entangled (R$^{max}_{sim}$), separable (R$^{sep}_{sim}$), {{and originating from annihilations in XAD-4 material (R$^{XAD-4}_{sim}$)}}.} 
    \begin{tabular}{|c|c|c|c|c|c|}
        \hline
         Range radius&Range centre& R$_{exp}$&R$^{max}_{sim}$&R$^{sep}_{sim}$& {R$^{XAD-4}_{sim}$}\\
         \hline
         r = 20$^\circ$ & $\theta_1 = \theta_2 =82^\circ$&2.00	$\pm$ 0.03 & 2.44 & 1.54 & {2.08}\\
         r = 20$^\circ$ & $\theta_1 = \theta_2=94^\circ$ & 1.93 $\pm$ 0.03 & 2.15 & 1.46 & {1.97}\\
         \hline
    \end{tabular}
     
    \label{table:Results}
\end{table}

In metals, positrons {exclusively} annihilate directly with electrons~\cite{Bass:2023rmp}, whereas in the porous polymer XAD-4 used in this study, positrons annihilate directly in only 32\% of cases and in {the remaining} 68\% of cases, the annihilation proceeds through the formation of positronium atoms~\cite{Jasinska:2016}. 
In this study, the conversion of orthopositronium on oxygen molecules is suppressed because air has been pumped out of the XAD-4 material to 10$^{-4}$~Pa using the dedicated chamber and vacuum system~\cite{Gorgol:2020acta}.
We hypothesize that the non-maximal entanglement can be attributed to the annihilation of positrons with electrons bound to the molecules, when a positron from positronium annihilates with an electron from the surrounding atoms
~\cite{Coffman:2000,Koashi:2004,Horodecki:2009}. 
In general, when annihilation is from the mixed state the entanglement may be partially or fully lost~\cite{Beatrix:2017,Caradonna:2024}. 
{
Using eq. 7, and taking into account that in the case of XAD-4 material, the pick-off process constitutes 31\% of 2$\gamma$ annihilations (see Table II) we obtain R(XAD-4) = 2.36. Pink squares in Fig.~3c and Fig.~3d show prediction for measured value R simulated using the value of R(XAD-4)~=~2.36, taking into account the properties of the detection system and the data selection criteria. The result obtained under the assumption that photons from pick-off process are not entangled (pink squares) is in quite good agreement with the experimentally determined values of R (blue circles). In Table I the calculated and measured values are compared for r~=~20$^\circ$.
}
To further explore the origin of the observed non-maximal entanglement, dedicated experiments are required in which the annihilation mechanism can be identified, e.g., by additional measurements of the positron lifetime. 
The observation reported in this work, based on measurements in the XAD-4 porous polymer, demonstrates that the degree of entanglement of annihilation photons, expressed via the R parameter, is not maximal.
This observation can be explained by assuming that photons from the pick-off annihilations are not entangled. The rate of the pick-off process in matter is determined by the size of the pores (free voids between atoms) in which positronium atoms are formed. The smaller the size of the free voids, the larger the pick-off contribution. It is well established that the larger the pick-off rate, the smaller the positronium lifetime~\cite{Bass:2023rmp}. Here, we anticipate that the R-value will also be smaller. The oPs lifetime is known to vary with tissue type, and it has been argued that the observed lifetime changes are predominantly due to the differences in the molecular tissue structure~\cite{Bass:2023rmp}. Therefore, the result of this work opens perspectives to apply R as a new diagnostic indicator that may be available in PET imaging.

\begin{table}[t!]
   {
    \centering
    \caption{{\bf{The fraction of the main processes leading to annihilation of positrons in XAD-4 materials~\cite{Jasinska:2016}}}}
    \label{tab:annihilation-processes}
    \begin{tabular}{|c|c|c|c|c|}
        \hline
        \textbf{Process}           & \textbf{Direct ($e^+e^- \to 2\gamma$)} & \textbf{$pPs \to 2\gamma$} & \textbf{$oPs \to 2\gamma$ (pick-off)} & \textbf{$oPs \to 3\gamma$} \\
        \hline
        \textbf{Total Fraction [\%]}  & 32                                  & 17                           & 22                                    & 29                         \\
        \textbf{Fraction of $2\gamma$ [\%]} & 45                                  & 24                           & 31                                    & 0                          \\
        \hline
    \end{tabular}
   }
\end{table}
The non-maximal entanglement of annihilation photons observed in this work, created in electron-positron annihilation in porous polymer, is well explained by the assumption that photons from the pick-off process are not entangled, while photons originating from direct and para-positronium annihilations are maximally entangled. However, the assumption that photons from direct annihilation are maximally entangled is based on previous measurements of the R-value in aluminum~\cite{Ivashkin:2023}, and the maximal entanglement of photons from para-positronium annihilation is predicted theoretically~\cite{Acin:2001}, but not yet confirmed experimentally. Thus, the main limitations of the studies presented in this article are the lack of R-value measurements for different materials with the same detector system and the lack of disentanglement between different annihilation mechanisms. Therefore, further systematic measurements with different materials are required to understand how variations in molecular structure affect the R-value. 
The parameter R has been used as an entanglement witness in annihilation photon studies to date. Recently, a strategy based on the fact that the analyzing powers and measure of correlations in the visibility function can be factorized was put forward in reference~\cite{Tkachev2025}, which allows for the extraction of these dependencies and thus the definition of a measure of entanglement degree independent of the scattering angle. The method proposed in reference~\cite{Tkachev2025} could reduce the required statistics and facilitate the comparison of results from different experiments. In the future, we intend to apply this approach to the J-PET detector.
Moreover, to unambiguously answer the question of how the $\Delta\varphi$ distribution (and hence the R-value) depends on the annihilation mechanism, the registration of prompt gamma, in addition to annihilation photons, will be required. Simultaneous registration of annihilation photons and prompt gamma will enable the determination of the positron lifetime in the studied material and hence will enable to disentangle annihilations proceeding directly, via para-positronium, or via pick-off processes.
The imaging of polarization correlations requires coincident registration of four interactions and therefore needs high-sensitivity scanners capable of multiphoton registration. New-generation PET scanners covering the whole body provide high enough sensitivity\cite{Spencer:2021,Prenosil:2022,Dai:2023}, yet for imaging of polarization correlation a multi-photon acquisition is required, which is not yet available in the current clinical PET systems. In the case of crystal PET systems, multi-pixel readout will also be necessary~\cite{Watts:2021,Parashari:2022,Ana:2021}. The first total-body PET scanner based on plastic scintillator is under construction at Jagiellonian University using J-PET technology~\cite{Moskal:2021pmb}. With the extended 250~cm field-of-view, this scanner will offer high multiphoton registration efficiency, enabling not only conventional 2$\gamma$ metabolic PET imaging with high statistical accuracy but also polarization correlation imaging and utilizing the recently developed positronium imaging ~\cite{Moskal:2021sra,moskal2024positronium} as a new diagnostic biomarker. 

%
%
\section*{{Methods}}\label{sec:methods}

\subsection*{\textbf{Positron annihilation chamber}}

In the reported experiment positrons were annihilating in the XAD-4 (CAS 37380-42-0) material\cite{Jasinska:2016}. Amberlite XAD-4 is a porous polymer resin with a high surface area and porous structure in which large fraction (68\%) of positron annihilations proceed via positronium formation~\cite{Jasinska:2016}. 
{In the XAD-4 polymer, positron annihilation proceeds in approximately 32\% via the direct $e^+e^- \to 2\gamma$ process, in approximately 17\% via $e^+e^- \to pPs \to 2\gamma$, in approximately 29\% via $e^+e^- \to oPs \to 3\gamma$, and in 22\% via $e^+e^- \to oPs$ followed by oPs pick-off annihilation into $2\gamma$. Thus, the two-photon annihilations proceed in 45\% via direct $e^+e^-$ annihilation, in about 24\% via self-annihilation of parapositronium, and in about 31\% via the pick-off process of orthopositronium (Table II).}
{The} $^{22}$Na isotope with an activity of 1.1~MBq was employed as a positron emitter. The $^{22}$Na source was sandwiched between two 7$\mu$m-thick kapton foils and enclosed on both sides by a few-mm-thick layer of the XAD-4 porous polymer. The polymer and the positron source were situated within a dedicated chamber connected to a vacuum system that enabled the evacuation of air from the polymer pores to a pressure of 10$^{-4}$ Pa~\cite{Gorgol:2020acta}. The chamber was placed at the center of the J-PET scanner as illustrated in Fig.~\ref{fig:2}(b). The chamber was constructed from 1~mm thick polyamide 6 (PA6) with a density of 1.14 g/cm$^3$, resulting in a negligible absorption (less than 1\%)
of 511 keV photons~\cite{Gorgol:2020acta}.
It is worth noting that a high vacuum was essential to suppress mechanisms that can be triggered by the presence of paramagnetic oxygen molecules, such as the conversion of oPs to pPs, which disrupt the oPs decay dynamics by decreasing the oPs lifetime, and also by decreasing the fraction of pick-off processes. However, vacuum is not required for quantum entanglement studies, and therefore, a similar methodology could be adapted for in vivo studies where differences in molecular composition can affect entanglement properties through different ratios of pick-off annihilations to direct and pPs annihilations.

\subsection*{\textbf{J-PET scanner}}\label{sec:meth:jpet}
J-PET is the first PET tomograph composed of plastic scintillators in which the Compton interaction is the main photon registration 
process~\cite{Moskal:2014nim}. The scanner used in this work consists of 192 plastic scintillators (EJ-230, $50 \times 1.9 \times 0.7$ cm$^3$) forming 3 concentric cylindrical layers (see Fig.~\ref{fig:2}(a))~\cite{Niedzwiecki:2017acta}.  
The scintillators are connected at the ends to the vacuum tube photomultipliers (Hamamatsu R9800). The time and position of the interaction of photons (referred to as hits) along the scintillators are estimated by measuring the time of light signal arrivals to the photomultipliers~\cite{Moskal:2014nim}. The signals from the photomultipliers are sampled at four fixed thresholds (in mV: 30, 80, 190, 300) using Time to Digital Converters (TDCs) implemented in Field Programmable Gate Arrays (FPGA) devices~\cite{Palka:2017wms}. The time stamps of the signals were recorded in the triggerless mode using the dedicated data acquisition system, which can process data streams at a speed of about 8 Gbps~\cite{Korcyl_ieee}. \\ 
The calibration procedure and intra-synchronization of the timing signals between 192 detection modules are explained in previous work. The hit-time and hit-position resolutions are equal to 250~ps and 25~mm, respectively~\cite{Niedzwiecki:2017acta}. {The maximum energy deposited by a 511 keV inside plastic scintillator corresponds to 340 keV, with an energy resolution of about 7.5\%}. The angular resolution for determining the scattering angle of annihilation photon amounts to 2$^\circ$~\cite{Moskal:2018}. 
\subsection*{\textbf{Events selection and classification}}
\begin{figure}
    \centering
    \includegraphics[scale = .12]{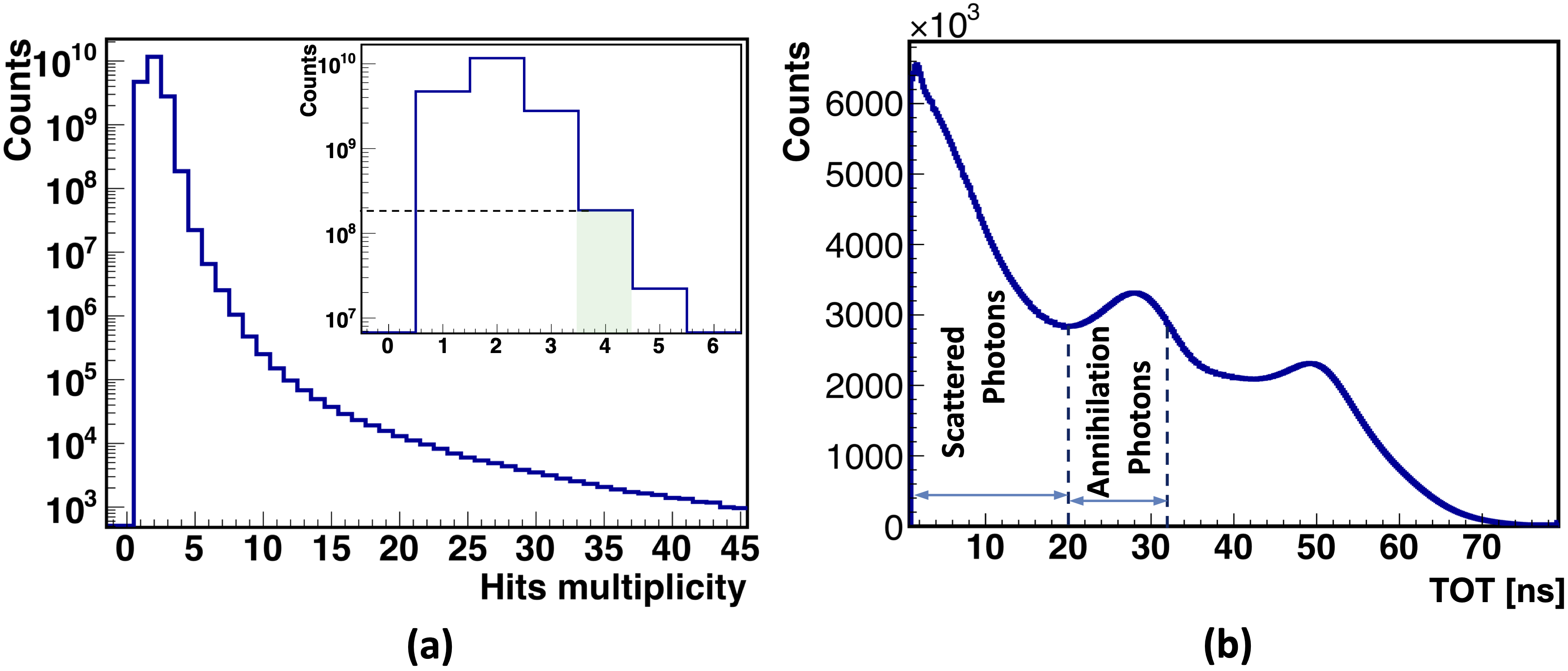}
       \caption{
    {     
        {\bf{Experimental distributions used in the data selection.}}   
    } 
    \textbf{(a)} Distribution of hit multiplicity in events. Inset highlights 4-hit events used for the analysis in this work. The number of events with multiplicity of 1 is suppressed by prescaling the data with single hits in the analysis. 
    \textbf{(b)} Time-over-threshold (TOT) histogram for all hits.
   Slopes at TOT values of about 32~ns and 56~ns correspond to the Compton edges resulting in maximum energy deposition by 511~keV annihilation photons, and 1275~keV prompt gamma (from $^{22}$Na decay), respectively. Dashed vertical lines indicate the range of TOT values used to select candidates for annihilation photons (20 - 32~ns), and scattered photons (1 - 20~ns). 
    \label{fig:5}
    }
\end{figure}

The data were collected continuously for 122 days and analyzed using a dedicated framework developed based on C++ and Root (a data analysis tool developed at CERN~\cite{ROOT:1997}) with detector-specific and advanced features.
Events useful for the study of the polarization correlations comprise four hits within 20 ns time window,  2 hits caused by primary photons (511 keV), and the remaining 2 hits caused by the corresponding scattered photons with lower energy (e.g. 275 keV for the scattering at 82$^\circ$). 
An example of such events is shown in Fig.~\ref{fig:2}(a) and Fig.~\ref{fig:2}(c). 
Fig.~\ref{fig:5}(a) shows the histogram of the hit multiplicity in the event. For the analysis only events with multiplicity equal to 4 were accepted (total events = 1,070,127,000). 
In the next step of analysis, as a first criterion to disentangle between the annihilation and scattered photons,
the TOT value was used which is a measure of the energy deposition~\cite{Sharma:2020}. 
For the 511 keV annihilation photons the energy depositions varies between 0 and 341 keV, and for the scattered photons of interest e.g. for the photons scattered at angles in the range of $82^\circ \pm 30^\circ$,
the maximum energy deposition is equal to 218~keV and 98~keV for photons scattered under 52$^\circ$ and 112$^\circ$, respectively.
Fig.~\ref{fig:5}(b) shows the distribution of TOT with superimposed vertical lines indicating ranges chosen for selecting scattered photons candidates (1~ns $<$ TOT $<$ 20~ns) and annihilation photons candidates (20~ns $<$ TOT $<$ 32~ns). 
The values of TOT are uniquely correlated with the energy deposition and, for example, TOT~=~32~ns corresponds to 341~keV (Compton edge for 511~keV photons), 
and TOT~=~56~ns corresponds to 1062~keV (Compton edge for 1275~keV gamma from $^{22}Na$ decay~\cite{Manish:2023}). 
Furthermore, for the annihilation photon candidates, the stringent back-to-back ($175^\circ <\theta <185^\circ$) emission criterion was applied. 

\begin{figure}[t]
    \centering
    \includegraphics[width = 0.95\textwidth]{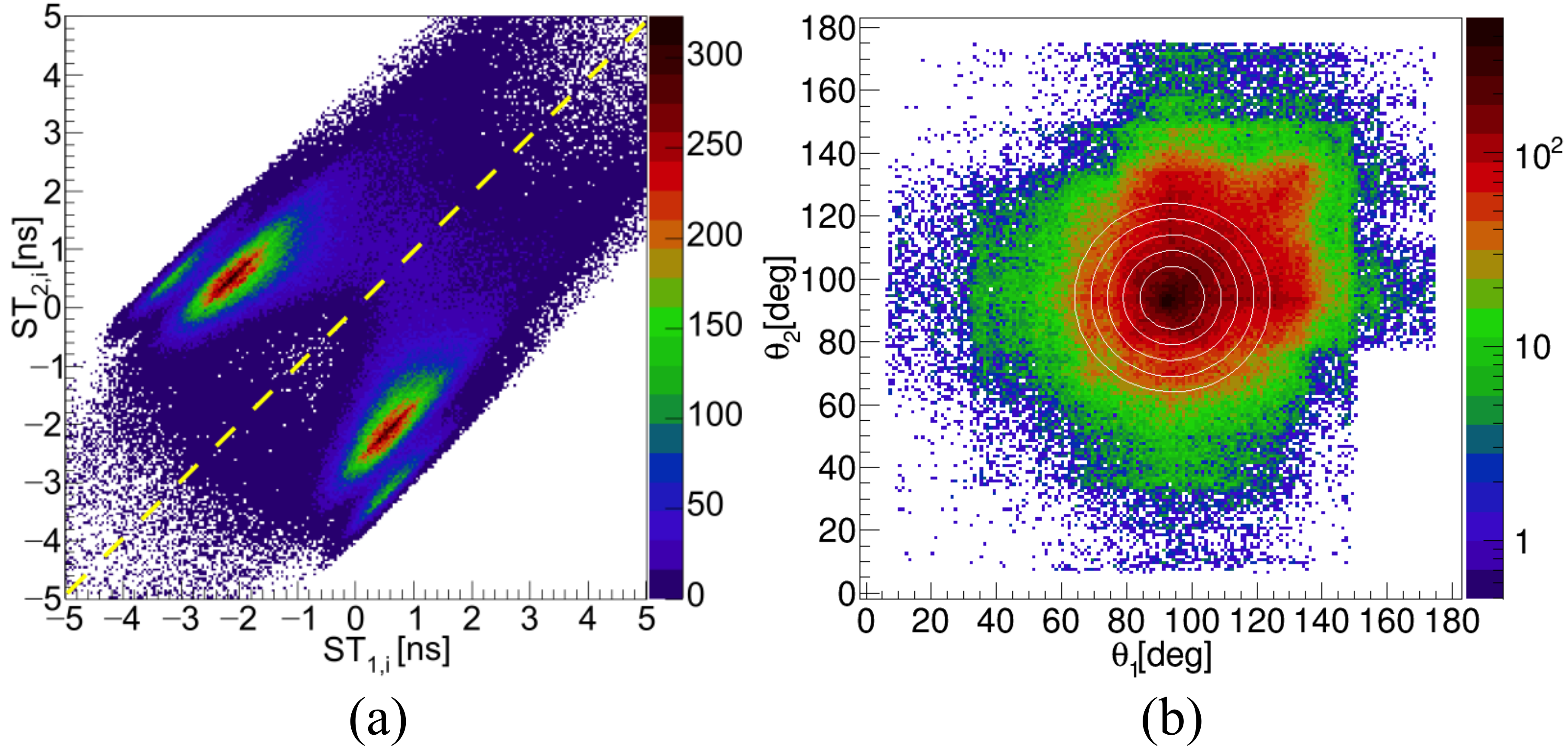}
    \caption{
    \textbf{Experimental distributions used for events classification.} \textbf{(a)} Distribution of ST$_{2i}$ vs. ST$_{1i}$ used to assign i$^{th}$ scattered photon to the 1$^{st}$ or 2$^{nd}$ annihilation photon. Diagonal dashed-yellow line is used as a criterion for the selection. If the (ST$_{1i}$,ST$_{2i}$) point is below the diagonal then it is assumed that the i$^{th}$ photon is a result of the scattering of the 1$^{st}$ annihilation photon. Contrary, if the (ST$_{1i}$,ST$_{2i}$) point is above the diagonal then i$^{th}$ scattered photon is assigned to the 2$^{nd}$ annihilation photon. The observed split structure is due to the arrangement of the scintillators (Fig.~\ref{fig:2}). \textbf{(b)} Experimental distribution of the $\theta_2$ vs. $\theta_1$ angles. Due to the geometrical arrangement of scintillators in the J-PET scanner (Fig.~\ref{fig:2}), the maximum density of the ($\theta_2$,$\theta_1$) distribution is for the scatterings around the angles of $\theta_1 = \theta_2 = 94^\circ$. The white circles indicate regions with the radii of r of $10^\circ, 15^\circ, 20^\circ, 25^\circ$ and $30^\circ$ used in the analysis (see Fig.~\ref{fig:3}).
    \label{fig:6}
    }
\end{figure}

Moreover, to ensure that photons are emitted from the source, an information from the tomographic image (Fig.~\ref{fig:2}(d)) was used. The image shows the density of annihilation points reconstructed based on the times and positions of hits identified as originating from annihilation photons.
For further analysis, only events were considered for which the distance between the reconstructed annihilation site
and the center of the image (position of the source) was less than 1~cm in the x-y plane and less than 4~cm in the z direction. 
After selecting the hits in the event corresponding to the annihilation photons, as a next step the hits due to the scattered photons are assigned to the proper annihilation photon.
By assigning indices 1 and 2 to the hits from annihilation photons and indices 3 and 4 to the hits from scattered photons, the next step in the analysis can be defined as testing two hypotheses: (i) the third photon is a result of scattering from the first annihilation photon and the fourth photon is a result of scattering from the second annihilation photon, and (ii) the third photon is a result of scattering from the second annihilation photon and the fourth photon is a result of scattering from the first annihilation photon. To test the hypothesis stating that i$^{th}$ photon is a scatter of the j$^{th}$ photon, we define a "Scatter Test" value: $ST_{j,i}$ = $\Delta t_{j,i} - r_{j,i}/c$, where $\Delta t_{j,i} = t_j^{annihilation} - t_i^{scatter}$ and $r_{i,j}$ is the distance between the hit positions of the annihilation photon($r_j$) and scatter photon($r_i$). ST is a difference between the measured time of flight of the scattered photon, and the time of flight calculated for the light to travel between the i$^{th}$ and j$^{th}$ hits.  In the case of ideal time and position resolution of the scanner, if the i$^{th}$ photon is a scatter of the j$^{th}$ photon than ST must be equal to zero. 
%
%
Fig.~\ref{fig:6}(a) shows the distribution of values ST$_{1,i}$ vs. ST$_{2,i}$ for testing the hypotheses whether the i$^{th}$ scatter photon originates from 1$^{st}$ or 2$^{nd}$ annihilation photon. If the point (ST$_{2,i}$,ST$_{1,i}$) on the plot is below the diagonal then the i$^{th}$ scatter photon is assigned to the 1$^{st}$ annihilation photon. If it is above the diagonal, then it is assigned to the 2$^{nd}$ annihilation photon.
For the final analysis, only these events were selected for which each of the two different scattered photons was assigned to a different annihilation photon.
After applying the sequential selection criteria to isolate events of interest, the number of remaining events was reduced (relative to the total number of initial events) to 92\% following the axial restriction on the active length of scintillators, 19\% after selecting annihilation and scattered photons based on the time-over-threshold (TOT), 2\% for selecting events containing exactly two annihilation photons and two scattered photons, 0.1\% after reconstructing the annihilation coordinates and ensuring that annihilation events originated from the source chamber, and 0.05\% after applying scattering test to correctly pair each scattered photon with its corresponding primary photon.
%
Finally, for each event the values of scattering angles $\theta_{1}$ and $\theta_{2}$, and the relative angle between the scattering planes ($\Delta\varphi$) are determined.
Fig.~\ref{fig:6}(b) shows the distribution of scatterings angle $\theta_{2}$ versus $\theta_{1}$. The maximum density of registered events is observed for $\theta_{2} = \theta_{1} = 94^\circ$. This is due to the geometrical arrangement of scintillators in the J-PET scanner, as can be seen in~Fig.\ref{fig:2}(a) and~Fig.\ref{fig:2}(c). The white circles superimposed on the experimental distribution in  Fig.~\ref{fig:6}(b) illustrate angular regions centered at $\theta_{2} = \theta_{1} = 94^\circ$ for which the $\Delta\varphi$ distributions and R parameter were determined and presented in Fig.~\ref{fig:3}(d).
\subsection*{\textbf{Estimation of J-PET scanner acceptance and registration efficiency}}
The correction of the measured distributions for the geometrical acceptance and registration efficiency of the J-PET scanner is one of the most crucial steps in determining the polarization correlation of annihilation photons. Efficiency and acceptance were simultaneously determined in this work, and for the sake of brevity, we will refer to this combined quantity as 'efficiency' throughout the manuscript.  Thus, in the following we will understand by efficiency the combined geometric acceptance, detection efficiency and event selection efficiency. Such defined efficiency was estimated as a function of the $\Delta\varphi$ angle using the GEANT4 simulation package~\cite{Geant4:2016}.
For this purpose, the full scanner geometry and material composition~\cite{Niedzwiecki:2017acta} were defined in the GEANT4 structure and the $e^+e^-$ annihilations were simulated in the center of the detector assuming the fraction of direct annihilations, pPs and oPs formations, as it is known for the XAD-4 porous polymer~\cite{Jasinska:2016}. The method of simulations with the J-PET scanner was validated in the previous works, as described e.g. in Ref~\cite{Moskal:2018,Moskal:2021}.
The response of the scanner was simulated. Next the simulated data were analysed using the same criteria as applied to the experimental data.
Then, such determined $\Delta\varphi$ distributions were normalized to the original $\Delta\varphi$ distributions determined by simulating the annihilation photons interactions in the scanner and taking for the calculations a true scatterings angles~\cite{Moskal:2018}. The resulting normalized $\Delta\varphi$ spectra were used to correct the experimental $\Delta\varphi$ spectra for efficiency. Example of corrected $\Delta\varphi$ spectra are shown in Fig.~\ref{fig:3}(c) and  Fig.~\ref{fig:3}(d).
\subsection*{\textbf{Simulations of $\Delta\varphi$ distributions for separable and entangled annihilation photons}}
 For simulations performed in this work it was assumed that the polarizations of the back-to-back propagating annihilation photons are perpendicular to each other. For the separable state interaction of each photon was simulated independently using \textsc{GEANT4} package, in which the Compton scattering is simulated according to the Klein-Nishina formula~\cite{KleinNishina:1929}. In order to simulate the distribution for entangled photons we used the data simulated for separable state and preselected events such that the resulting distributions are as expected for the entangled photons. The method used for simulations of separable and quantum entangled photons was described in detail, and validated in Ref.~\cite{Moskal:2018}.
\subsection*{\textbf{Systematic uncertainties}}
 The precision of the R-value achieved in this study, at the level of 10$^{-2}$, is primarily limited by statistical uncertainties, while the systematics of the detector performance, as already demonstrated in previous works, is well controlled up to the level of at least 10$^{-4}$~\cite{Moskal:2018,Moskal:2021pmb,Sharma:2020,Moskal:2021,Moskal:2024}. The measurement was purposely performed with a very low activity source of 1.1~MBq, resulting in accidental coincidences of less than~2\%. 
 However, this was achieved at the cost of a very long measurement time of 122 days. 
 A signal event is comprised of four hits registered in four different scintillator strips 
 (2$\gamma$ + 2$\gamma^\prime$). The most important feature of the detector is that, in a given layer, each scintillator contributes equally to the registration of all configurations. Therefore, even if the efficiency of a given scintillator were not well estimated, only the statistics would be decreased; the shape of the $\Delta\varphi$ distribution, however, would not be changed. And only the shape is important for the results of these studies. In total, 192 scintillator strips with an angular distance of 1.875$^\circ$ are located at an average radius of $\sim$49.5~cm. In a given layer, each single strip contributes equally to the final result. For the sake of argument, the full removal of a single detector (1 out of 192) would have a 5 × 10$^{-3}$ effect on the statistics, whereas an exaggerated misplacement of a strip by 0.1~cm would have a 0.14$^\circ$ effect, while the angular coverage of a single strip in the XY plane is 0.5$^\circ$. So, even if the efficiency and geometry of a given scintillator were determined with a precision of 10$^{-2}$, the effect on the total efficiency would be at the level of 10$^{-5}$. However, the total efficiency does not influence the accuracy of the R determination. Here, the crucial factor is the relative efficiency between the registration of various $\Delta\varphi$ angles. And since each scintillator contributes to the measurement of all configurations, the R determination is not affected by inaccuracies in the efficiency determination of single detector strips. Additionally, with the possibility of image reconstruction, the average position of the source can be controlled within 0.5~mm in the XY plane and 0.4~mm along the Z axis, while the thickness of the scintillators is 7~mm. The influence on the $\Delta\varphi$ distribution and R extraction of the effects of primary and secondary photon identification, assignment of secondary to primary photons, and, in general, the influence of all criteria used in the data selection process were tested by changing these criteria and reanalyzing the data. In all cases, no statistically significant change in the result was found. It is worth mentioning that the full geometry of the J-PET detector, including all the aluminum elements of the mechanical construction for possible scattered events, was simulated. No influence of such events was found. Figures 3c and 3d show that the averaging of R over the angles, simulated and measured, follows the same pattern, and that the conclusion of this article does not change when the range of $\theta_1$ vs. $\theta_2$ angles is changed by using a different range of scattering angles.



%
\bibliography{Non-maximalENT} 
\bibliographystyle{sciencemag}


\section*{Acknowledgments}
The authors would like to thank Professor Steven Bass for helpful discussions.
\paragraph*{Funding:}
 We also acknowledge the support provided by 
the Foundation for Polish Science through the 
TEAM POIR.04.04.00-00-4204/17 Program (P.M.); \\
National Science Centre of Poland through grants no.
2019/35/B/ST2/03562 (P.M.),\\ 2021/42/A/ST2/00423 (P.M.), \\ 2021/43/B/ST2/02150~(P.M.), \\2022/47/I/NZ7/03112~(E.Ł.S), \\
2020/38/E/ST2/00112~(E.P.d.R.),\\ 2022/06/X/ST2/01444~(S.S.); \\
the Ministry of Education and Science through grants 
no. \\ SPUB/SP/490528/2021 (P.M.),\\ IAL/SP/596235/2023 (P.M.);\\
 as well as the SciMat and qLife Priority Research Area budgets under the programme Excellence Initiative - Research University at the Jagiellonian University (P.M. and E.Ł.S.).
\paragraph*{Author contributions:}
The experiment was conducted using the Jagiellonian Positron Emission Tomograph (J-PET). The J-PET scanner, the techniques of the experiment and this study was conceived by P.M. The data analysis was conducted by D.K. Signal selection criteria were
developed by P.M. and S.S., applied by D. K., and verified by S.S.
Authors: P.M., D.K., S.S, E.Y.B., S.C., N.C., A.C., C.C., E.C., M.D., K.D., M.G., B.J., K. Kacprzak, T. Kaplanoglu, Ł.K.,
T. Kozik, 
E.L., F.L., W.M., S.N., S.P., E.P.d.R.,  M.R., M.S., E.Ł.S., P.T., K.T.A., and K.V.E., and participated in the construction, commissioning, and operation of the experimental setup, as well as in the data-taking campaign and data interpretation. N.C. under the supervision of S.N. and P.M. monitored the whole data taking campaign. M.G. and B.J. designed and constructed the positronium production chamber. S.N. optimized the working parameters of the detector. K.D. and K. Kacprzak, took part in developing the J-PET analysis and simulation framework. K.D., M.S., K.K. and E.P. del Rio performed timing calibration of the detector. E.C. developed and operated short- and long-term data archiving systems and the computer center of J-PET. S.S. established relation between energy loss and TOT and dependence of detection efficiency on energy deposition. 
S.P. performed simulations for sensitivity profiles of metabolic, entanglement and positronium imaging. 
P.M. managed the whole project and secured the main financing. E.Ł.S contributed to securing financing of the project. The results were interpreted by P.M., S.S. and D.K. The manuscript was prepared by P.M., S.S., and D.K. and was then edited and approved by all authors.
\paragraph*{Competing interests:}
There are no competing interests to declare.

\paragraph*{Data and materials availability:}
All data needed to evaluate the conclusions in the paper are present in the paper and/or the Supplementary Materials.
%
%


\end{document}